\documentclass[%
 reprint,
showpacs,
 amsmath,amssymb,
 aps,
]{revtex4-1}

\usepackage{graphicx}
\usepackage{dcolumn}
\usepackage{bm}

\begin{document}
\preprint{APS/123-QED}
\title{Dynamic analysis and continuous control of semiconductor lasers}
\author{Sohrab Behnia}
 \email{s.behnia@iaurmia.ac.ir}
\affiliation{Department of Physics, Faculty of Science, IAU, Orumia Branch, Ourmia, Iran.}%
\author{Khosro Mabhouti}
\affiliation{ Department of Physics, Faculty of Science, Ourmia university,Ourmia, Iran.}%
\author{Saeid Afrang}
\affiliation{ Department of Electrical Engineering, Ourmia MEMS lab, Ourmia university,Ourmia, Iran.}%
\date{\today}
\begin{abstract}
Stability control in laser is still an emerging field of research.
In this paper the dynamics of External cavity semiconductor lasers
(ECSLs) is widely studied applying the methods of chaos physics. The
stability is analyzed through plotting the Lyapunov exponent
spectra, bifurcation diagrams and time series. The results of the
study show that the rich nonlinear dynamics of the electric field
intensity ($\mid E \mid^2$) includes saddle node bifurcations,
periodic, quasi periodic, chaotic and regular pulse packages. The
issue of finding the conditions for creating stable domains has been
studied. By considering the dynamical pumping current system coupled
with laser, a method for the creation of the stable domain has been
introduced.
\begin{description}
\item[Usage]
 \end{description}
\end{abstract}
\maketitle
\section{Introduction}
The research on the laser dynamics has attracted a lot of interest
for the past several years. Current investigations in this field
include the study of different routes to chaos in lasers and the
control of chaos in such systems. Most of these works are
concentrated on semiconductor lasers with respect to their
importance in optical communications. Chaotic behavior in
semiconductor lasers has been investigated under various physical
conditions such as external optical injection,
and external optical feedback ~\cite{L1,L2,SC3}.\\
Our study is motivated from different points of view, in this study
dimensionless ECSLs are investigated~\cite{SC5}. This class of
dynamical systems can successfully be modeled by delay differential
equations. Where the study  has provided a realistic background for
the investigation of time delay systems, which is very difficult as
a problem in the infinite-dimensional space~\cite{DDE1}. This allows
us to present a consistent overall picture of the dynamics at the
same time. In this study parameters such as feedback phase, feedback
strength, and pump current are used as control parameters to obtain
different dynamical regimes.\\ Also the key idea underlying most
chaos controlling schemes such as ECSLs is to take advantage of the
unstable steady states and unstable periodic orbits of the
system~\cite{LC1,LC2,LC3}. They are embedded in the chaotic
attractor characterizing the dynamics in phase space. Feedback
schemes require real-time measurement of the state of the system and
processing of a feedback signal~\cite{FLC1,FLC2}. The location of
the unstable fixed-point must be determined before control is
initiated. The standard control method has the
disadvantage of having a small region of convergence around the fix point~\cite{C1,C2,C5}.\\
The aim of the present paper is studying the ECSLs dynamics by
the Lyapunov exponent diagram as the analyzing tool and
proposing a control method for the Laser. Regarding the pump current
as one of the control parameter of the system, a discrete
time dynamical system is built and coupled with Laser. \\
 The rest of the paper has been
organized as follows: Section 2 describes Lang and Kobayashi model
for ECSLs. The method of the control and analyzing the results are
proposed in Section 3 and 4.  In section 5, the stable domain of the
introduced model before and after applying the control method is
studied via Lypaunov exponents and bifurcation diagrams. Section 6
concludes the letter. The paper ends with an Appendix which contains
algebraic calculations of the controller.
\section{The Laser model}
In the early 1980s, Lang-Kobayashi (LK) proposed model semiconductor
lasers. The LK equations are model equations that have been used
extensively in the past to describe a semiconductor laser subject to
feedback from an external cavity~\cite{LK1,LK2}. LK equations are a
sort of delay differential equations (DDEs) with an
infinite-dimensional phase space~\cite{SC3}. For the (complex)
electric field E and inversion N, we write the LK equations as the
dimensionless and compact set of equations~\cite{SC5}:
\begin{equation}
\frac{dE}{dt}=(1+i\alpha)NE+\eta E(t-\tau)e^{-iC_p},
\end{equation}
\begin{equation}
T\frac{dN}{dt}=P-N-(1-2N)\mid E\mid^2.
\end{equation}
Parameters describe the line width enhancement factor $\alpha$, the
feedback strength $\eta$, the $2\pi$-periodic feedback phase $C_p$,
the ratio between carrier and photon lifetime $T$ and the pump
current $P$. In these equations, the time is normalized to the
cavity photon lifetime $(1 ps)$ ~\cite{Time}. The external round
trip time $\tau$ is also normalized to the photon lifetime. The
remaining parameters are held fixed at $T=1710,$ $P=0.8$, $\tau=70$,
$\alpha=5.0$ ~\cite{cte1}. In this paper, the stability of an
electric field intensity  $\mid E \mid^2$ is studied versus the
feedback phase change $C_p$ and feedback strength change $\eta$.
\section{The control method}
Most recently, a different method of control has been shown to be
successful in the experimental control of chaos. Based on:
\begin{itemize}
  \item Determination of the stable and unstable directions in the Poincare section.
  \item Self-controlling feedback procedure.
  \item Introduction of small modulation of a control parameter.
  \item Knowledge of a prescribed goal dynamics.
\end{itemize}
 The first two methods  are usually called feedback methods, while the 3rd and last are
called non-feedback methods~\cite{feedback,non-feedback}. Although control of chaos by
small modulations has not been proved in general~\cite{CONTROL_2},
this method involves the on chaotic behavior generation of an error
signal from the difference between the output signal and its value
at an earlier time. A great virtue of this method is that it does
not require knowledge of more than one variable. The method should
be very useful for applying in fast systems to control a chaotic
nonlinear circuit. Until now, different parameters such as
modulating current~\cite{CONTROL_1,CONTROL_5}, modulating
voltage~\cite{CONTROL_3}, and pump power variation response to a
continuous error signal generator~\cite{CONTROL_4} have been used to
modify the laser output. In this study, ``pump current'' has been
selected as a tool to be implemented in the control method. By
considering the Laser and the control system as a two dimensional
dynamical system, the simple model for controlling the stability of
Laser is introduced as follows~\cite{Beh}:
\begin{equation}
 \left\{
\begin{array}{l}
\dot{E}(E,N,P_{m})= \left\{
\begin{array}{l}
 \frac{dE}{dt}=(1+i\alpha)NE+\eta E(t-\tau)e^{-iC_p},\\\\
 T\frac{dN}{dt}=P_{m}-N-(1-2N)\mid E\mid^2,
     \end{array}\right.\\\\
P_{m+1}=\frac{4P_m}{(1-P_m)^2}.
      \end{array}\right.
 \end{equation}
The setup is shown schematically in Fig. 1. The most significant
parts of the setup include the laser pump and the method whereby its
output is modulated. The pump power is controlled by the use of an
electric circuit. Figure 2 shows the desired controller. A
photodiode is used to convert a sample of laser light to electric
voltage ``$V_s$''. This voltage is connected to non-inverting input
of the comparators ``$U_1$'' and ``$U_2$''. In the stable condition,
$V_s$ is less than $V_1$ and more than $V_2$. Thus, the output
voltage of $U_1$ and $U_2$ is in the low and high conditions,
respectively. Consequently, $Q_1$ and $Q_2$ are off and $Q_3$ is on.
As a result the relays 1 and 2 are off and $P_m$ is connected to the
input of the $P_{m+1}$ function generator. When the unstable
condition occurs and the laser intensity is increased, the voltage
$V_s$ becomes more than $V_1$ and then the output voltage of $U_1$
changes to high. Due to high condition in the output of $U_1$, the
transistor $Q_1$ and $Q_2$ conduct. At this stage the relay 2
disconnects $P_m$ from the input of the $P_{m+1}$ function generator
and simultaneously relay 1 connects the $P_{m+1}$ function generator
output to its input. The feedback from the output to the $P_{m+1}$
function generator input continuously decreases the $P_{m+1}$, and
consequently decreases photodiode output voltage. Finally, when the
output voltage of the photodiode becomes less than $V_1$, $Q_1$ and
$Q_2$ become off. In this condition relay 1 disconnects but relay 2
is still on. And the capacitor ``$C$'' at the input of $P_{m+1}$
function generator holds the last $P_{m+1}$. After a while, if laser
intensity decreases more, and ``$V_s$'' becomes less than $V_2$,
then, the output voltage of $U_2$ changes to zero and then $Q_3$
changes to off. As a result, relay 2 becomes off and connects the
$P_m$ to the input of the $P_{m+1}$ function generator. Finally,
$P_{m+1}$ increases and the system changes to normal condition.\\ In
this study, arbitrary chaotic regions of the system dynamical
behavior, have been chosen to undergo the proposed control
method~\cite{CONTROL_2,CONTROL_1,CONTROL_5}.
\section{Stability analysis}
\subsection{Lyapunov exponent spectrum}
Lyapunov exponents and entropy measures, can be considered as
``dynamic'' measures of attractors complexity and are called ``time
average''~\cite{Bif1}. The Lyapunov exponent $\lambda$ is useful for
distinguishing various orbits. Lyapunov Exponents quantify
sensitivity of the system to initial conditions and give a measure
of predictability. The Lyapunov exponents are a measure of the rate
at which the trajectories separate one from another. A negative
exponent implies that the orbits approach to a common fixed point. A
zero exponent means that the orbits maintain their relative
positions; they are on a stable attractor. Finally, a positive
exponent implies that the orbits are on a chaotic attractor, so the
presence of a positive Lyapunov
exponent indicates chaos. The Lyapunov exponents are defined as follows:\\
Consider two nearest neighboring points in phase space at time $0$
and $t$, with the distances of the points in the $ith$ direction
$\|{\delta}x_i(0)\|$ and $\|{\delta}x_i(t)\|$, respectively. The
Lyapunov exponent is then defined by the average growth rate
$\lambda_i$ of the initial distance,
\begin{equation}
{\lambda_i}=
\lim_{t{\rightarrow}{\infty}}{\frac{1}{t}}\ln{\frac{\|{\delta}x_i(t)\|}{\|{\delta}x_i(0)\|}}.
\end{equation}
The existence of a positive lyapunov exponent is the indicator of
chaos showing neighboring points with infinitesimal differences at
the initial state abruptly separate from each other in the $ith$
direction~\cite{Lya1}. Using the algorithm of Wolf \cite{Wolf}, the
lyapunov exponent is calculated versus a given control parameter.
Then, the value of the control parameter increases a little and the
Lyapunov exponent is calculated for the new control parameter. By
continuing this procedure Lyapunov exponent spectrum of the system
is plotted versus the control parameter.
\subsection{Bifurcation diagrams}
 Bifurcation means a qualitative change in the dynamical behavior of a system when a
parameter of the system is varied~\cite{Bif1}. A bifurcation diagram
provides a useful insight into the transition between different
types of motion that can occur as one parameter of the system
alters. It enables one to study the behavior of the system on a wide
range of an interested control parameter. In this paper the
dynamical behavior of the system is studied through plotting the
bifurcation diagrams of the of intensity $\mid E \mid^2$ versus
feedback phase $C_p$ and feedback strength change $\eta$, as control
parameters. This procedure continued by increasing the control
parameter and the new resulting points were plotted in the
bifurcation diagram versus the new control parameter.
\section{Results and Discussions}
\subsection{Introducing the dynamics of the master laser}
The dynamical behavior of the master laser as a function of the
control parameter, $C_p$, can be divided into two areas:
\begin{itemize}
  \item low feedback strengths
  \item high feedback strengths
\end{itemize}
 In the low feedback strengths, such as; $\eta_m$= 0.0455~\cite{Time}, the output
of laser contains chaotic (CH) ~\cite{Chaos}, periodic $(P1,P2,..)$
and quasi periodic (QP) ~\cite{QP}, behaviors, respectively. Fig.
3(a) shows that the dynamical behavior of the output intensity for
master laser is CH, P7, P1 and QP. In the high feedback strengths,
such as; $\eta_m$= 0.135, the laser output contains regular pulse
package (RPP), P1, and QP behaviors~\cite{Time,cte1}, which is
confirmed in  Fig. 4. As it is shown, the dynamical behavior of
 the output intensity for master laser is RPP, QP and P1.
For understanding the general faces of the ECSLs one could refer to
Fig. 5 , where the bifurcation curve of the laser output in terms of
 feedback strength $\eta$, in a constant feedback phase $C_p$ is presented. As it is shown, the bifurcation curve
 contains periodic, QP, CH, and RPP
 behaviors.\\
 The maximum Lyapunov exponents are presented in Fig. 3(b) to verify the
corresponding characteristics. As can be observed, the positive and
negative values in Lyapunov exponent spectrum verify the aperiodic
and periodic behaviors in bifurcation diagram. In addition, extended
numerical analysis on the ECSLs dynamics has been carried out on
several control parameters. The system exhibits extremely nonlinear
behaviors when the control parameters are varied.
\subsection{Applying the chaos control method}
In order to provide a fair justification on employing a periodic
perturbation as the control technique, it is first necessary to
consider two facts. First, the similar chaos control methods should
be evaluated, and their practicability and associated advantages
should be clarified. Second, we address this question that: can such
a method be easily implemented in the real world applications? In
fact, a method should be chosen so as to be best suited for an
experimental implementation. In this way, as it will be discussed in
the following, the pump current
sources will be able to satisfy both highlighted remarks.\\
Before evaluating the potential chaos control methods for an ECSLs
system, we should consider the parameters that can be implemented in
the control technique. It is clear that, to control the chaotic
oscillations of a dynamical system, one of the control parameters of
the system should be perturbed or modified in order to reach the
regular desirable behavior. However, in the application phase , we
are restricted to apply changes only to the external forcing
elements such as feedback phase, feedback strength that are
determined by the type of the application, and the pump current which has been chosen in the present study.\\
The results are depicted in Figs. 6-9. Under the collective system
parameters, every category corresponds to original system and
controlled system; see Figs. 6(a), 6(b) and Figs. 7(a), 7(b). Also
the control method has been tested through Lyapunov exponent
diagrams; see Figs. 6(c) and 7(c). The dashed line represents the
original system, while the solid line represents the controlled
system. This figure indicates a significant abatement of the
Lyapunov exponents from positive values to negative ones indicating
that stable dynamics can be achieved after the proposed technique is
engaged. The results have also been confirmed by plotting the
intensity $\mid E \mid^2$ versus time in a certain value of the
feedback phase $C_p$ before and after the control process.
Respectively Figs. 8(a) and 9(a) show before control process and
Figs. 8(b) and 9(b) depict after control process.
\section{Conclusion}
Our main contribution in this paper is to develop the simple method
for study and control the laser dynamics. In this paper, the concept of feedback chaos control
 has been re-defined by taking into account the control parameter as a variable in the time
 and
  is changed by using another chaotic map, for which a new and effective control scheme has been
presented. The result of the present study broadens our
understanding of the complex dynamics of laser, and also helps us to
control its nonlinear dynamics.
\begin{acknowledgments}
The author expresses his sincere thanks to Professor A. Jafari for
his invaluable advice and encouragement.
\end{acknowledgments}
\appendix
\setcounter{equation}{0}
\section{Detail of control circuit}
The circuits in the Figs. 10, 11 and 12 are used to generate
$P_{m+1}$ Function generator. Figure 11 shows the multiplier
circuit. This circuit multiplies $(1-P_m)$ by $(1-P_m)$ and makes
the denominator of the $P_{m+1}$ function generator. The derived
equations for Fig. 11 are as follows:
$$V_{b}=-V_D= -V_TLn(\frac{I_D}{I_S}+1)$$
\begin{equation}
=-V_TLn(\frac{V_a}{RI_S}+1),
\end{equation}
The op amp $U_5$ adds the two equal voltages $V_b$ to its inverting
input. Then, the voltage $V_c$ at the output $U_5$ is:
$$
V_c=V_T\left(Ln(\frac{V_a}{RI_S}+1)+Ln(\frac{V_a}{RI_S}+1)\right)$$
\begin{equation}
=V_TLn\left((\frac{V_a}{RI_S}+1)^2\right),
\end{equation}
and the voltage at the output of $U_6$ is:
$$V_d=-RI_{D1}=-RI_s\left(e^{(\frac{V_c}{V_T})}-1\right)$$
\begin{equation}
=-\frac{V_a^2}{RI_s}-2V_a,
\end{equation}
The op amp $U_7$ adds the voltages $V_d$ and $V_a$ , and therefore,
cancels ``-2$V_a$'' at its output. Then, the voltage at $V_g$ will
be as follows:
\begin{equation}
V_g=\frac{V_a^2}{RI_s},
\end{equation}
By considering $R=1M \Omega$ and $I_s=1 \mu sec$, the equation
$(A4)$ can be written as:
\begin{equation}
V_g=V_a^2=(1-P_m)^2,
\end{equation}
In Fig. 12, a part of multiplier circuit including op amp, $U_2$,
$U_4$, $U_5$ and $U_6$ is used as a divider. In this figure, the
voltages at points a and b are respectively  $4P_m$ and $-(1-P_m)^2$
, therefore, the voltage at point c is as below:
$$
V_c=V_T\left(Ln(\frac{4P_m}{RI_S}+1)-Ln(\frac{(1-P_m)^2}{RI_S}+1)\right)$$
\begin{equation}
=V_TLn\left(\frac{4P_m+RI_s}{(1-P_m^2)+RI_s}\right),
\end{equation}
Assuming $4P_m\gg RI_s$ and $(1-P_m)^2\gg RI_s$, the equation $(A6)$
can be written as below:
\begin{equation}
V_c= V_T Ln\left(\frac{4P_m}{(1-P_m)^2}\right),
\end{equation}
and then $V_d$ is:
\begin{equation}
V_d = -RI_s (e^{\frac{V_c}{V_T}}-1)\cong -RI_s\frac{4P_m}{(1-P_m)^2},
\end{equation}
Finally $V_f$ is:
\begin{equation}
V_f = \frac{4P_m}{(1-P_m)^2}.
\end{equation}
\appendix

\bibliography{apssamp}
\end{document}